\documentclass[12pt]{article}
\usepackage{amsmath}
\usepackage{graphicx}
\usepackage{enumerate}
\usepackage{natbib}
\usepackage{url} % not crucial - just used below for the URL 

\usepackage{bm}
\usepackage{amssymb}

\newcommand{\indep}{\perp \!\!\! \perp}
\DeclareMathOperator*{\argmax}{arg\,max}

\newtheorem{theorem}{Theorem}[section]
\usepackage{algorithm}
\usepackage[noend]{algpseudocode}
\usepackage{amssymb}
\usepackage{amsfonts}
\usepackage{mathrsfs}
\usepackage{multirow}
\usepackage{booktabs}
\usepackage{verbatim}
\usepackage{array}
\usepackage{caption}
\usepackage{setspace}
\usepackage{fixltx2e}% for \textsubscript
\usepackage[compact]{titlesec}
\usepackage{siunitx} %align numbers by decimal point
\newcommand{\blind}{1}

\begin{document}

\def\spacingset#1{\renewcommand{\baselinestretch}%
{#1}\small\normalsize} \spacingset{1}

%%%%%%%%%%%%%%%%%%%%%%%%%%%%%%%%%%%%%%%%%%%%%%%%%%%%%%%%%%%%%%%%%%%%%%%%%%%%%%

\if1\blind
{
  \title{\bf Non-greedy Tree-based Learning for Estimating Global Optimal Dynamic Treatment Decision Rules with Continuous Treatment Dosage}
  \author{Chang Wang, Lu Wang\\
    Department of Biostatistics, University of Michigan}
  \maketitle
} \fi

\if0\blind
{
  \bigskip
  \bigskip
  \bigskip
  \begin{center}
    {\bf Non-greedy Tree-based Learning for Estimating Global Optimal Dynamic Treatment Decision Rules with Continuous Treatment Dosage}
\end{center}
  \medskip
} \fi

\bigskip
\begin{abstract}
Dynamic treatment regime (DTR) plays a critical role in precision medicine when assigning patient-specific treatments at multiple stages and optimizing a long term clinical outcome. However, most of existing work about DTRs have been focused on categorical treatment scenarios, instead of continuous treatment options. Also, the performances of regular black-box machine learning methods and regular tree learning methods are lack of interpretability and global optimality respectively. In this paper, we propose a non-greedy global optimization method for dose search, namely Global Optimal Dosage Tree-based learning method (GoDoTree), which combines a robust estimation of the counterfactual outcome mean with an interpretable and non-greedy decision tree for estimating the global optimal dynamic dosage treatment regime in a multiple-stage setting. GoDoTree-Learning recursively estimates how the counterfactual outcome mean depends on a continuous treatment dosage using doubly robust estimators at each stage, and optimizes the stage-specific decision tree in a non-greedy way. We conduct simulation studies to evaluate the finite sample performance of the proposed method and apply it to a real data application for optimal warfarin dose finding.
\end{abstract}

\noindent%
\vfill

\newpage
\spacingset{1.9} % DON'T change the spacing!

\section{Introduction}
\label{sec:intro}

A dynamic treatment regime (DTR) is a sequence of decision rules that determine the optimal treatment for individual patients at multiple stages. Treatment decisions are based on each patient's unique characteristics and medical history to optimize their long-term clinical outcomes. With the emergence of precision health care, DTRs with continuous dosage treatment have become increasingly important, allowing for personalized optimal dosage intervention. Examples include optimal dose finding in radiation oncology therapy and drug trials with multiple stages.

Although black box learning methods like random forest and deep learning can produce accurate predictions of the optimal treatments, their lack of interpretability makes them difficult for medical experts to understand and implement. Conventional tree learning methods such as CART are easy to interpret and predict, but they often use greedy algorithms that can fail to converge to the global optimum or achieve high performance under certain circumstances. Additionally, while most research on dynamic treatment regimes has focused on selecting the optimal treatment type, there is limited knowledge on DTRs with continuous dose finding. Another challenge is how to maintain desirable properties like doubly robustness and asymptotic normality for continuous dose finding in DTR settings.

The literature on evaluating dynamic treatment regimes is extensive and can be divided into two categories for finding the optimal tree-based DTR. One approach is to use supervised learning methods, which involve estimating the optimal treatment first and then transforming it into tree versions (known as the treatment-tree algorithm). Several algorithms are commonly used for optimal treatment estimation, including Q-learning (e.g. \citealp{Qlearning}), Marginal Structural Models (e.g. \citealp{MSM}), and Outcome Weighted Learning (e.g. \citealp{OWL}). However, most of these methods require a correct specification of working models for the propensity model or the conditional outcome mean model, which can be challenging to validate with limited knowledge. Therefore, a method with doubly robustness, which can ensure consistent estimation as long as one of two models is correctly specified, is desirable.

 The other tree-based learning approach is searching for trees with the best performance, also known as creating a new tree and evaluating its performance under the optimal treatment (denoted as the tree-treatment algorithm). However, this approach often relies on either random/stochastic search, which lacks efficiency (e.g. \citealp{STRL}), or greedy search, which may converge to a local optimum and fail to find the global optimum as it only maximizes the current purity measure (e.g. \citealp{TRL}). In other words, both approaches have limitations, with the treatment-tree algorithm relying on stochastic or greedy search and most treatment-tree algorithms not being doubly robust. Therefore, we aim to propose a treatment-tree algorithm with doubly robustness that avoids stochastic/greedy tree learning and applies a non-greedy algorithm.

The second challenge arises when developing an individualized causal inference algorithm for continuous treatment dosage with doubly robustness. The majority of published papers about DTRs deal with categorical treatment assignment. While there are some algorithms that can estimate doubly robust causal effects and find optimal continuous treatment at the population level, these methods require the use of a tree-treatment algorithm, which necessitates the decision of the study population before the algorithms can be applied. As a result, stochastic or greedy searching cannot be avoided. Alternatively, some methods can estimate the optimal continuous treatment at the individual level by optimizing a well-designed objective function. However, these methods are not robust as some depend on the correct specification of the conditional outcome model, while others rely on the correct specification of the propensity model \citep{LZ,CZK}. Therefore, we aim to propose a new method for causal inference that can achieve individualized optimal dose with doubly robustness.

To overcome the aforementioned challenges, we introduce a new method called Global Optimal Dosage Tree-based learning (GoDoTree), which is a non-greedy tree-based optimization technique that estimates optimal DTRs in a multi-stage continuous-treatment environment, using data from randomized trials or observational studies to provide personalized intervention with patient-specific medicine dosage. At each stage, GoDoTree constructs a decision tree by first modeling an individual-level counterfactual treatment effect curve via semiparametric regression models, and then performing a non-greedy tree-based search to optimize the counterfactual treatment effect. The two main components of GoDoTree, namely individual-level counterfactual outcome estimation and tree learning, are implemented in a backward manner at every stage to optimize the long-term objective function. The proposed GoDoTree has several advantages, including great interpretability, doubly robustness, and the ability to achieve global optimal DTRs. It also contributes to the existing literature on the development of DTRs with continuous treatment. To demonstrate the performance of GoDoTree, we conduct simulation studies to show its global optimization ability and apply it to warfarin dosing data collected from \citealp{warfarin}.

This paper is structured as follows. Section 2 introduces the proposed method for individualized counterfactual outcome estimation. In Section 3, we describe non-greedy tree-based supervised learning, which allows for a global optimal tree search with known counterfactual outcome for every sample. The details of finding the optimal kernel function are presented in Section 4. Section 5 formalizes the problem of estimating optimal DTRs with continuous treatment and outlines the framework of our algorithm. In Sections 6 and 7, we present the results of numerical studies and an application example, respectively. Finally, we provide a concluding discussion in Section 8.

\section{Individualized Counterfactual Outcome Estimation With Continuous Dosage}

We will firstly introduce treatment regime optimization with single stage. In single stage scenario, we denote $Z_i = (\bm{X}_i,A_i,Y_i)$ as the observed data for patient $i$, where $\bm{X}_i$ is a vector of covariates, $A_i$ is a continuous treatment or exposure, and $Y_i$ is the outcome of interest. $Y^a$ is the counterfactual outcome when patient takes the treatment $a$ and $\mu(\bm{x},a) = \mathbb{E}[Y^a \mid \bm{X} = \bm{x}]$ is the conditional expectation of $Y^a$.

Population level effect curve $\theta_p(a) = \mathbb{E} Y^{a} =\int_{\mathscr{X}} \mu(\bm{x},a) dP(\bm{x}) $ is frequently used in causal inference with continuous treatment, which is the potential outcome that would have been observed under continuous treatment level $A=a$ for the whole population. However, to perform global tree optimization, we define a new individual-level effect curve $\theta_i(a)$ for the $i$ th patient as
 
$$\theta_i(a) =
    \frac{\int_{\mathscr{X}} \mu(\bm{x},a)K_{i}(\bm{x})  dP(\bm{x})}{\int_{\mathscr{X}} K_{i}( \bm{x})  dP(\bm{x})} =  \frac{ \mathbb{E} \mu(\bm{X},a)K_{i}(\bm{X}) }{\mathbb{E} K_{i}( \bm{X})},$$
where $K_{i}(\bm{x})$ is a pre-specified kernel function centered at $\bm{X}_i$ and gives samples different weights according to their similarities with sample $i$. In another word, the neighborhood of sample $i$ are detected by $K_i$ and this subgroup's weighted population level effect curves are defined as the individual level effect curve for sample $i$. 

Instead of optimizing the population level outcome, we propose a surrogate objective function and optimize the individual level potential outcome: $g^{opt} = \argmax \ \sum_{i=1}^{n} \theta_{i}( g(\bm{X}) )$.

\begin{algorithm}
\caption{Individual level effect curve estimation}
\begin{algorithmic}[1]
    \State  Fit a conditional outcome mean model of $\mu(\bm{x},a)$ and a propensity model of $\pi(a\lvert \bm{x})$. Get estimates $\hat{\mu}(\bm{x},a)$ and $\hat{\pi}(a\lvert \bm{x})$, with $\sup \lvert\hat{\pi}- \pi^{*}\rvert = o_p(1) $ and $\sup \lvert \hat{\mu}- \mu^{*}\rvert = o_p(1)$.

% $\lVert\hat{\pi}- \pi^{*}\rVert_{\mathcal{Z}} = o_p(1) $, $\lVert\hat{\mu}- \mu^{*}\rVert_{\mathcal{Z}} = o_p(1)$.

    \State In order to estimate the effect curve $\theta_i(a)$, we propose a doubly robust estimator $\hat{\xi}(\bm{Z})$: 
$$\hat{\xi}(\bm{Z}{;}\hat{\pi},\hat{\mu},\hat{\kappa}) = \hat{\kappa}^{-1} \left( \frac{Y-\hat{\mu}(\bm{X},A)}{\hat{\pi}(A\mid \bm{X})}\hat{w}(A)K_{i}(\bm{X}) + \hat{m}(A) \right)$$
where 
$\hat{w}(a) = \int_{\mathcal{X}} \hat{\pi}(a\lvert \bm{x}) d\mathbb{P}_n(\bm{x}), \hat{m}(a) = \int_{\mathcal{X}} K_{i}(\bm{x}) \hat{\mu}(\bm{x},a) d\mathbb{P}_n(\bm{x}), \hat{\kappa} = \int_{\mathcal{X}} K_{i}(\bm{x}) d\mathbb{P}_n(\bm{x}).$
    \State Estimate patient-specific effect curve $\hat{\theta}_i(a)$ for every patient $i$:
$$\hat{\theta}_i(a) = \gamma_{ba}(a)^T \hat{\bm{D}}^{-1}_{ba} \mathbb{P}_n\{ \gamma_{ba}(A)  K_{ba}(A) \hat{\xi}(\bm{Z}{;}\hat{\pi},\hat{\mu},\hat{\kappa}) \},  $$
where $\hat{\bm{D}}^{-1}_{ba} = \mathbb{P}_n\{ \gamma_{ba}(a)  K_{ba}(A) \gamma_{ba}(a)^T\}$, $\gamma_{ba}(A) = (1,(A-a)/b)^T$, $K_{ba}(A)$ is a kernel function centered at $a$ and scaled with $b$.

 % \Comment{root to leave}

\end{algorithmic}
\end{algorithm}

\subsection{Working Model Estimation}

The estimations in the conditional outcome mean model and the propensity model are vital for estimating optimal DTRs. Researchers can construct parametric models with subject-relevant knowledge or just use some nonparametric models like random forest, generalized additive model. Specifically, Bayesian additive regression trees (BART) values as a great candidate because it requires minimal specification of the model and can approximate complex functions well with a tree ensemble \citep{BART}. Nonparametric Kernel Smoothing Methods are also frequently used to fit propensity model \citep{np}.

\subsection{Doubly Robust Estimator of the Individualized Effect Curve}

%Since we will estimate effect curves $\theta_i(a), i=1,2,...,N$， for all samples, we will simplify $\theta_i(a)$ to $\theta(a)$ when it causes no confusions. 

We simplify the notation of $\theta_i(a)$ to $\theta(a)$ when it causes no confusions, as we estimate effect curves $\theta_i(a), i=1,2,...,N$ for all samples. To derive doubly robust estimators for $\theta(a)$, we adapt semiparametric theory in a novel way similar to the approach of \citealp{Rubin2005} and \citealp{Kennedy_2016}. Our goal is to find a function $\xi(\bm{Z}{;}\pi,\mu,\kappa)$ of the observed data $\bm{Z}$ and nuisance functions $(\pi,\mu)$ with doubly robustness, which means $\mathbb{E}\{ \xi(\bm{Z}{;}\pi^*,\mu^*,\kappa) \lvert A=a\} = \theta(a)$ if either $\pi^* = \pi$ or $\mu^* = \mu$ (no necessarily both). Here $\kappa = \mathbb{E} K(\bm{X})$ is a constant when the kernel function $K$ is fixed.

According to the semiparametric theory, a doubly robust mapping is related to the efficient influence function for a certain parameter. If $\mathbb{E}\{ \xi(\bm{Z}{;}\pi^*,\mu^*,\kappa) \lvert A=a\} = \theta(a)$, then it follows $\mathbb{E}\{ \xi(\bm{Z}{;}\pi^*,\mu^*,\kappa)\} = \psi$ for $\psi = \mathbb{E} \theta(a)  =   \int_{\mathscr{A}} \theta(a)w(a) da$, where $w(a)$ is the marginal probability density function of treatment $A$. This indicates that a component of the efficient influence function for the parameter $\psi$ can be a candidate for the doubly robust mapping $\xi(\bm{Z}{;}\pi,\mu,\kappa)$\citep{robins2001, vanderlaan2003}. We derive the efficient influence function for $\psi$ as follows:

\begin{theorem}[Efficient Influence Function] 

Under a semiparametric model, $\xi - \psi + (\mathbb{E} K_i(\bm{X}))^{-1} \int_{\mathscr{A}}[\mu(\bm{X},a)K_i(\bm{X}) - m(a)]w(a)da$ is the efficient influence function for $\psi = \int_{\mathscr{A}} \theta_i(a)w(a) da$, where $w(a) = \int_{\mathcal{X}}\pi(a\lvert x)dP(x), m(a) = \int_{\mathcal{X}}\mu(x,a ) K_i(\bm{x}) dP(x)$ and

%Under a nonparametric model, the efficient influence function for $\psi = \int_{\mathscr{A}} \theta_i(a)w(a) da$ is $\xi - \psi + (\mathbb{E} K_i(\bm{X}))^{-1} \int_{\mathscr{A}} [\mu(\bm{X},a)K_i(\bm{X}) - m(a)]w(a)da$, with

$$\xi(\bm{Z}{;}\pi,\mu) = (\mathbb{E} K_i(\bm{X}))^{-1} [\frac{Y-\mu(\bm{X},A)}{\pi(A\mid \bm{X})}w(A)K_i(\bm{X}) + m(A)].$$

%$$w(a) = \int_{\mathcal{X}}\pi(a\lvert x)dP(x) = \mathbb{E} \pi(a\lvert \bm{X}), m(a) = \int_{\mathcal{X}}\mu(x,a ) K_i(\bm{x}) dP(x) = \mathbb{E} \mu(\bm{X},a ) K_i(\bm{X}) $$

\end{theorem}

Based on this, we propose a new estimator $\xi$, which is a component of the efficient influence function of $\psi$, and further prove its doubly robustness.

\begin{theorem}[Doubly Robustness]
The proposed estimator $\xi$ has the doubly robust property: if $\pi^{*}=\pi$ or $\mu^{*} = \mu$, then we have
$$\mathbb{E} \{\xi(\bm{Z}{;}\pi^{*},\mu^{*},\kappa)\lvert A=a\}  = \theta(a).$$
\end{theorem}

So as long as one of conditional outcome mean model $\mu$ or propensity model $\pi$ is correctly specified (not necessarily both), our estimator $\xi$ will be unbiased for $\theta(a)$.

\subsection{Individual Level Effect Curve Estimation}

With the doubly robust mapping $\xi(\bm{Z}{;}\pi,\mu,\kappa)$ derived in the previous subsection, for which $\theta_i(a) = \mathbb{E}\{ \xi(\bm{Z}{;}\pi^{*},\mu^{*},\kappa) \lvert A=a\}$ as long as $\pi^* = \pi$ or $\mu^* = \mu$, we can construct doubly robust estimate $\hat{\xi}(\bm{Z}{;}\hat{\pi},\hat{\mu},\hat{\kappa})$ and regress on treatment variable $A$. The local linear kernel version of the estimator is $\hat{\theta}_b(a) = \gamma_{ba}(a)\hat{\beta}_b(a)$, where 
$$\hat{\beta}_b(a) =  \operatorname*{arg\,max}_{\beta\in\mathbb{R}^2} \mathbb{P}_n [K_{ba}(A)\{\hat{\xi}(\bm{Z}{;}\hat{\pi},\hat{\mu},\hat{\kappa}) - \gamma_{ba}(a)^T\beta \}^2], \gamma_{ba}(A) = (1,(A-a)/b)^T,$$
for $K_{ba}(t) = b^{-1} K\{ (t-a)/b \}$, with K a standard kernel function and $b$ a scalar bandwidth parameter. Then we have the close form for estimation of $\theta_i(a)$:
$$\hat{\theta}_i(a) = \gamma_{ba}(a)^T \hat{\bm{D}}^{-1}_{ba} \mathbb{P}_n\{ \gamma_{ba}(A)  K_{ba}(A) \hat{\xi}(\bm{Z}{;}\hat{\pi},\hat{\mu},\hat{\kappa}) \},  $$
where $\hat{\bm{D}}^{-1}_{ba} = \mathbb{P}_n\{ \gamma_{ba}(a)  K_{ba}(A) \gamma_{ba}(a)^T\}$.

We will show the asymptotic bias and normality of the proposed individual level effect curve $\hat{\theta}(a)$:

\begin{theorem}[Asymptotic Bias Analysis]

Let $\pi^{*}$ and $\mu^{*}$ denote fixed functions to which $\hat{\pi}$ and $\hat{\mu}$ converge in the sense that $\sup\lvert\hat{\pi}- \pi^{*}\rvert = o_p(1) $ and $\sup\lvert\hat{\mu}- \mu^{*}\rvert = o_p(1)$, and let $a \in \mathcal{A}$ denote a point in the interior of the compact support $\mathcal{A}$ of treatment $A$. Assume $[ \mu(\bm{X},t) - \hat{\mu}(\bm{X},t)][ \pi(t\lvert\bm{X}) - \hat{\pi}(t\lvert\bm{X}) ] = O_p(r_n)$, along with several regularity assumptions (see Appendix), we have:

(a) Either $\pi^{*}=\pi$ or $\mu^{*}=\mu$, where $\mu$ and $\pi$ are the true conditional outcome model and propensity model respectively.

(b) The bandwidth $b=b_n$ satisfies $b\to0$ and $nb^3 \to \infty$ as $n\to \infty$.

(c) $K$ is a continuous symmetric probability density.

(d) $\theta(a)$ is twice continuously differentiable, and both $\pi(a)$ and the conditional density of $\xi(Z{;} \pi, \mu)$ given $A = a$ are continuous as functions of $a$.

(e) The estimators $(\hat{\pi}, \hat{\mu},\hat{\kappa})$ and their limits $(\pi, \mu,\kappa)$ are contained in uniformly bounded function classes with finite uniform entropy integrals (as defined in Section 4 of the Appendix), with $1/\hat{\pi}$, $1/\pi$, $1/\hat{\kappa}$ and $1/\kappa$  also uniformly bounded.

%$$\hat{\xi}(\bm{Z}{;}\hat{\pi},\hat{\mu},\hat{\kappa}) = \hat{\kappa}^{-1} \left( \frac{Y-\hat{\mu}(\bm{X},A)}{\hat{\pi}(A\mid \bm{X})}\hat{w}(A)K(\bm{X}) + \hat{m}(A) \right)$$

%$$\hat{w}(a) = \int_{\mathcal{X}} \hat{\pi}(a\lvert \bm{x}) d \mathbb{P}_n(\bm{x}), 
%\hat{m}(a) = \int_{\mathcal{X}} K(\bm{x}) \hat{\mu}(\bm{x},a) d\mathbb{P}_n(\bm{x}), \hat{\kappa} = \int_{\mathcal{X}} K(\bm{x}) d\mathbb{P}_n(\bm{x})$$ 

%The local linear kernal version of the estimator is $\hat{\theta}_h(a) = \bm{g}_{ha}(a)\hat{{\beta}}_h(a)$, where $\bm{g}_{ha}(a) = (1,(t-a)/h)^T$ and 

%$$\hat{{\beta}}_h(a) =  \operatorname*{arg\,max}_{\beta\in\mathbb{R}^2} \mathbb{P}_n [K_{ha}(A)\{\hat{\xi}(\bm{Z}{;}\hat{\pi},\hat{\mu},\hat{\kappa}) - \bm{g}_{ha}(a)^T\beta \}^2]$$

$$\hat{\theta}_b(a) - \theta(a) = O_p(\frac{1}{\sqrt{nb}} + b^2  + r_n).$$

%\begin{align*}
    %&\hat{\theta}_b(a) = \bm{g}_{ba}(a)^T \hat{\bm{D}}^{-1}_{ba} \mathbb{P}_n\{ \bm{g}_{ba}(A)  K_{ba}(A) \hat{\xi}(\bm{Z}{;}\hat{\pi},\hat{\mu},\hat{\kappa}) \}\\
    
%\end{align*}

\end{theorem}

Here $O_p(r_n)$ denotes the convergence rate of conditional outcome model and propensity model. Since our first assumption is one of two models are correctly specified, without loss of generality, let us assume the conditional outcome mean model is correctly specified while propensity model not, i.e., $\pi^{*} \not =\pi$ and $\mu^{*}=\mu$. Then $\hat{\mu}-\mu = \hat{\mu}-\mu^* = O_p(r_n)$ and  $\hat{\pi}-\pi = O_p(1)$, because the latter one is biased but still bounded. Thus $( \mu - \hat{\mu})( \pi - \hat{\pi} ) = O_p(r_n)$ represent the convergence rate of the correctly specified model, when one of them might be biased.

In the next theorem we show that when one or both of $\hat{\pi}$ and $\hat{\mu}$ are estimated with fast enough convergence, then the proposed estimator is asymptotically normal after scaling.

\begin{theorem}[Asymptotic Normality]

Along with the same assumptions in theorem 2.3, also assume the convergence rate $r_n$ satisfies $r_n = o_p(1/\sqrt{nb})$, then we have

$$ \sqrt{nb} \{\hat{\theta}_b(a) - \theta(a) + bias(a) \} \overset{d}{\to} N(0,\frac{\sigma^2(a) \int K^2(u)du}{w(a)}), $$
where $bias(a) = \theta''(a)(b^2/2)\int u^2K(u)du + o(b^2)$, and 
$$
\begin{aligned}
\sigma^2(a) &= \mathbb{E} \{ [\xi(\bm{Z}{;}\pi^{*},\mu^{*},\kappa)- \theta(a)]^2  \mid A=a \} \\
&= \kappa^{-2} \mathbb{E}\left[ \frac{var \{ Y \lvert \bm{X},A=a\} + \{ \mu(\bm{X},a) -  \mu^{*}(\bm{X},a) \}^2}{\{ \pi^{*}(a\lvert\bm{X}) / w^{*}(a) \}^2 / \{ \pi(a\lvert\bm{X}) / w(a) \} }  
\right] - \{\theta(a) - \kappa^{-1}m^{*}(a)\}^2. \end{aligned}
$$
\end{theorem}

%Although the convergence rate is $1/\sqrt{nb}$, the bandwidth $b$ is not the larger the better because there exists a bias term. 

With all assumptions satisfied, we can see that the mean square error of $\hat{\theta}$ is $O(1/nb + b^4)$ and the variance-bias trade-off determines the theoretical optimal bandwidth $b \sim n^{-1/5}$. As for data-driven bandwidth selection, we treated $\hat{\xi}$ as known and used leave-one-out cross-validation bandwidth selection \citep{bandwidth}:

$$\hat{b}_{opt} =  \argmax_{b} \sum_{i=1}^n \left\{ \frac{\hat{\xi}(\bm{Z}_i{;}\hat{\pi},\hat{\mu}) - \hat{\theta}_b(A_i) }{1-\hat{W}_b(A_i)}  \right\}^2,$$
where $\hat{W}_b(a_i) = (1,0)\mathbb{P}_n\{\bm{\gamma}_{ba_i}(A)K_{ba_i}(A)\bm{\gamma}^T_{ba_i}(A) \}^{-1}(1,0)^Tb^{-1}K(0)$ is the $i^{th}$ diagonal element of the hat matrix.

 \label{chapter2}

\section{Global Optimal Tree Search Algorithm} 
%\subsection{Notation}

%In the last section, we have known every sample's pseudo outcome estimation 
%$\Tilde{PO}_{it}(a_t) := \hat{E}(\Tilde{Y}_t\mid A_t = a_t, \bm{H}_{it})$, $i \in \{1,2,...,N\}$, $a_t \in 
%\mathscr{A}$ 
%and our question can be reduced to fit an optimal and interpretable %continuous treatment regime $g_t: \bm{H}_{t} \to A_t$, $t \in \{1,2,...,T\}$, %to maximize the objective function:

%\begin{equation*}
%   L_t(g_t) = \mathbb{P}_n \hat{E}(\Tilde{Y}_t\mid A_t = g_t(\bm{H}_t), %\bm{H}_t) = 
%   \sum_{i=1}^{N} \Tilde{PO}_{it}( g_t )
%\end{equation*}
%
%where $\Tilde{PO}_{it}(g_t) := \hat{E}(\Tilde{Y}_t\mid A_t = g_t(\bm{H}_{it}), %\bm{H}_{it})$ is the estimated pseudo outcome at stage $t$, when patient $i$ %follows the treatment regime $g_t$.

%For simplicity, in the following sections we only introduce the above single stage scenario. It can be easily extended to multiple stage scenarios by replacing $\bm{X}$ with $\bm{H}_t$ at stage $t$. The outcome to be optimized is still the effect curves estimated in section 3, regardless it is regressed on single stage $Y$ or multiple stage $\Tilde{PO}_{it}(a_t)= \hat{E}(\Tilde{Y}_t\mid A_t = a_t, \bm{H}_{it})$.

We use non-greedy tree-based learning to search for a global optimal decision tree. Following last section's algorithm, the individual level outcome $\theta_i(a)$, $i \in \{1,2,...,N\}$, $a \in \mathscr{A}$, is estimated with $\hat{\theta}_i(a)$ as the input of supervised learning. In this section, we will simplify $\hat{\theta}_i(a)$ to $\theta_i(a)$ for brevity, because it is treated as the known ground truth. Then the question can be formalized to fit an optimal and interpretable continuous treatment decision tree $g: \bm{X} \to A$, to maximize the objective function: $L(g)  = \sum_{i=1}^{N} \theta_i(g(\bm{X}))$.

%(Decision tree is a tree-based algorithm where paths from root to leaf represent classification rules. Take binary tree for example, where each node has one parent and two children, each node represents a single input variable x and a split rule on that variable.
%: for numeric variables, the splitting rule will be ``greater than threshold $\tau$ or not''{;} for categorical variables, the splitting rule will be ``belong to set $S$ or not''. And the leaf nodes of the tree contain an output to make a prediction. It is noteworthy that all internal node (non-leaf) don't make any prediction, but partition the samples into several smaller parts. And only the leaf nodes make predictions based on samples assigned to them. Eventually, every sample will travel through the branches and reach a leaf node, and its predicted value will be given by that leaf node.)

Regular tree regression algorithms such as CART are developed in a greedy manner. They grow from top-down without backtracking, determining all input variables and split rules locally to optimize the current purity function. However, greedy algorithms have potential limitations such as failing to update parent nodes once they have a child node and the fact that the cumulative local optimal rules do not necessarily lead to a global optimal rule. Greedy algorithms are likely to miss stronger splits if they are hidden behind weaker ones, resulting in sub-optimal or overly complicated trees \citep{greedy1}. To overcome these limitations, we use a non-greedy algorithm called Tree Alternating Optimization \citep{TAO} to search for the global optimum with iterative optimization. Given a decision tree, we can optimize any node and keep the rest of the tree structure unchanged. By alternating optimization over the depth levels of the tree, the decision tree can optimize its structure, escape from local optima, and is more likely to converge to the global optimum than if it were to keep optimizing based on sub-optimal prior nodes. Figure 1 shows a comparison between greedy trees and non-greedy trees.

\paragraph{Optimization at internal node}
%\subsubsection*{Optimization at internal node}
Consider an internal node (non-leaf) and the corresponding subset of samples $\mathscr{S}$, our target is to find the optimal partition rule $\eta$ which divides $\mathscr{S}$ into two parts $\omega$ and $\mathscr{S} \backslash \omega$. 

 It is important to note that each element in $\mathscr{S}$ will eventually be assigned to one of the children, and both children and their descendants are fixed decision trees denoted as $g^{left}$ and $g^{right}$. If $\bm{X}$ is classified to the left child, it will be assigned to treatment $g^{left}(\bm{X})$, otherwise, it will be assigned to treatment $g^{right}(\bm{X})$. The optimization of $\eta$ can be reduced to a supervised classification problem where we seek the optimal rule $\eta$ that maximizes the surrogate objective function and partitions the current sample subset $\mathscr{S}$ into two parts $\omega$ and $\mathscr{S} \backslash \omega$:
\begin{equation*}
    \max_{\eta} W(\eta) = \max_{\eta} \sum_{\substack{ \bm{X}_i \in \omega }}   \theta_i\{g^{left}(\bm{X}_i)\} +
    \sum_{\substack{ \bm{X}_i \in \mathscr{S} \backslash \omega}}  \theta_i\{g^{right}(\bm{X}_i)\}.
\end{equation*}

The node update rule in GoDoTree is based on the idea of simulated annealing, which allows us to avoid getting trapped at a local optimum by introducing a probabilistic element into the decision-making process. Specifically, let $\eta_j, j=1,2,\dots,p$, denote the optimal splitting rule if the $j$ th variable is fixed as the classifier to maximize $W(\eta)$ for the current node, then the node update rule is 
$P(\eta = \eta_j) =   exp\{  \alpha_t W(\eta_j)\} /\sum_{k=1}^p exp\{  \alpha_t W(\eta_k)\}$, where $\alpha_t$ is a increasing sequence about the iteration number $t$.

 \begin{figure}[H]
\caption{Comparison between greedy tree and non-greedy tree algorithms. Greedy tree optimizes the current objective function and cannot update nodes once they are constructed, while the proposed approach, Tree Alternating Optimization (TAO), is a non-greedy tree search that can update any node in the tree at any time, with other nodes fixed. TAO can alternate optimization over depth levels of the tree and escape from local optima. The classifier used for partitioning is denoted as $\delta$.}
\centering
 \includegraphics[scale=0.5]{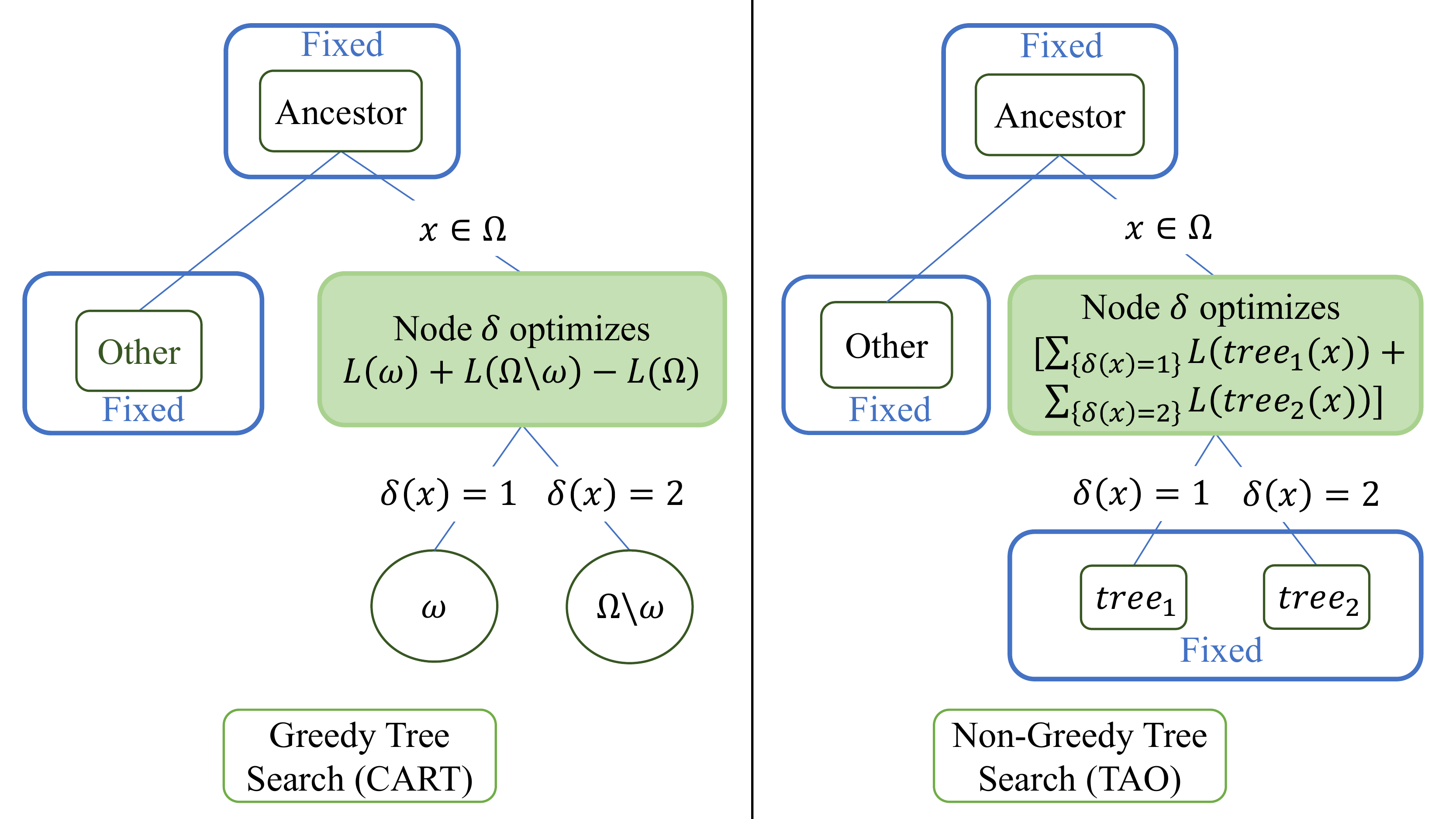}
\end{figure}

% (suroggate)
% two theta
% cite name

\paragraph{Optimization at leaf node}
%\subsubsection*{Optimization at leaf node}
As for the prediction part, our target is to optimize the parameter $\eta = a \in \mathscr{A}$ so as to maximize the surrogate objective function
$W(\eta) =  \sum_{  \bm{X}_i \in \mathscr{S}}    \theta_i(\eta)$. Samples classified to the same leaf node will be assigned to the same dose $\eta$.

%\subsubsection*{TAO algorithm framework}
\paragraph{TAO algorithm framework}
After updating the rule for each node, global optimization of the tree can be achieved through alternating optimization over the depth levels of the tree. The depth levels are cycled in the order root-leaf-root until convergence, with the criteria for convergence being either a numerically converging objective function or a fixed tree topology for several iterations. Pruning is not performed until convergence, and when the decision tree converges, any leaf node with a size smaller than a pre-specified number $n_0$ will be collapsed with its sibling nodes.

\begin{algorithm}[tbh]
\caption{Tree alternating optimization}
\begin{algorithmic}[1]
    \State Initialize a decision tree with height h, set t=0.
    \While{not converge} 
    \For{\texttt{i in 1 to h}}    \Comment{root to leave}
        \For{\texttt{every node in height i}}
        \State Find the optimal splitting rule $\eta_k$ when k th variable is used as predictor.
        \State Find current loss function W($\eta_k$).
        \State Update node parameter $\eta$ with probability $P(\eta = \eta_j) = \frac{   exp(  \alpha_t W(\eta_j))    }{\sum_{k=1}^p exp(  \alpha_t W(\eta_k))}$.
        \EndFor
    \EndFor
    \For{\texttt{every leaf node}}
        \State Find the optimal treatment.
    \EndFor
    
    \For{\texttt{i in h to 1}}    \Comment{leave to root}
        \State Do the same as row 4-7.
    \EndFor
    \State t = t+1
    
    \EndWhile\label{euclidendwhile}

\end{algorithmic}
\end{algorithm}

\section{Implementation: Search For Optimal Kernel} 
\begin{algorithm}
\caption{Search for optimal kernel}
\begin{algorithmic}
\State Step 1: Initialize $\hat{\theta}_i(a) = \hat{\mu}(x_i,a)$ using BART (Bayesian additive regression tree).

\State Step 2: Define distance matrix $D(i,j) = \sup_{a} \hat{\theta}_i(a) + \sup_{a} \hat{\theta}_j(a) - \sup_{a} \{\hat{\theta}_i(a) + \hat{\theta}_j(a)\}$.

\State Step 3: Calculate the similarity matrix $S_{ij} = cor(d_i,d_j)$, where $d_i = (d_{i1},d_{i2},...,d_{in})$ is the $i$ th row of the distance matrix $D$. 

\State Step 4: Calculate another similarity matrix $\Tilde{S}_{ij}$, using weighted Euclidean distance $\Tilde{D}$, where $\Tilde{D}_{ij} = \sum_{k=1}^{p} w_k(x_{ik} - x_{jk})^2$, $w_k$ is the variable importance estimated in Step 1.  

\State Step 5: Define new kernels as $K_{i}(\textbf{x}_j) = exp\{ \min(S_{ij},\Tilde{S}_{ij})/\sigma_i^2 \}$, where $\sigma_i^2$ is selected such that $\sum_{j=1}^{n} K_{i}(\textbf{x}_j) \approx n_{leaf}$, the expected sample number assigned to tree's leaves (e.g. $n/8$). 

%\State Step 5: With given kernels, estimate the individual level effect curves $\hat{\theta}_i$ for all samples ($i=1,2,...,n$), as we introduced in section 3.
%\State Step 6: Iterate step 3-5 until the clustering result converges.
\end{algorithmic}
\end{algorithm}

The idea of individualized effect curve estimation is, for a specific sample, find its neighborhood by using a kernel and giving different weights, and study the individualized effect curve based on this weighted sub-population. Thus the key point is to find an optimal kernel function. An ill-defined kernel will make the estimation of individual-level effect $\theta_i(a)$ seriously biased from true value.

%Also, if we simply $K_{l}(\textbf{x}) = K( \lVert x - l \rVert^2  /\sigma^2 )$ with $K$ a kernel function, it may fail to find correct neighborhood when information signal-noise rate is low. 

% The ``neighborhood'' definition should be based on their effect curves, while we need to use ``neighborhood''/kernel to calculate effect curve, which seems a cycle. So we decide to estimate effect curves and create kernels simultaneously with iterations.
 
 We propose a new distance measurement to evaluate the similarity of effect curves. For continuous treatment, the similarity between two samples should be based on their effect curves, rather than simply optimal treatment doses. Our newly proposed distance $D$ has two useful properties. Firstly, the distance is relevant to horizontal translation, but immutable to vertical translation. Secondly, smooth curves are close to any other curves under our definition. A smooth effect curve means that the outcome is not strongly associated with the treatments, thus this sample can be assigned to any treatments and regarded as many samples' neighborhood. So, a variable will not influence the distance if it only has main effect but no interaction with treatment. $\Tilde{D}$ is weighted to get rid of low information signal-noise rate. Because we only care about the optimal dose, the variable importance $w_k$ here is not about the main effect, but the interaction between variable and treatment. Finally, we incorporate two similarities with $\min(S_{ij},\Tilde{S}_{ij})$ to search for neighborhood with both effect curve and covariate similarities. Users are also allowed to use other similarity definitions, e.g. $\max(S_{ij},\Tilde{S}_{ij})$, which will search for neighborhood with either effect curve or covariate similarities.

To illustrate the distance definition, we demonstrate four example effect curves and their distances(See Figure 2 for visualization). The effect curves of A and B have the same shape but different horizontal location/optimal dose, thus their distance is large, which is denoted by the vertical line. A and C have different shapes and locations, but C is a smooth curve. So we can regard C as anyone's neighborhood and distance between A and C is relatively small. A and D have different vertical location but the same shape, and their distance is zero.

 \begin{figure}[H]
\caption{New distance visualization with $D(i,j) = \sup_{a} \theta_i(a) + \sup_{a} \theta_j(a) - \sup_{a} \{\theta_i(a) + \theta_j(a)\}$. Four effect curves are used as examples in the left figure. Three curves in the right figure represent the component $\theta_i(a) + \theta_j(a)$ and the vertical lines in the right figure visualize the pairwise distance measure.}
\centering
 \includegraphics[scale=0.7]{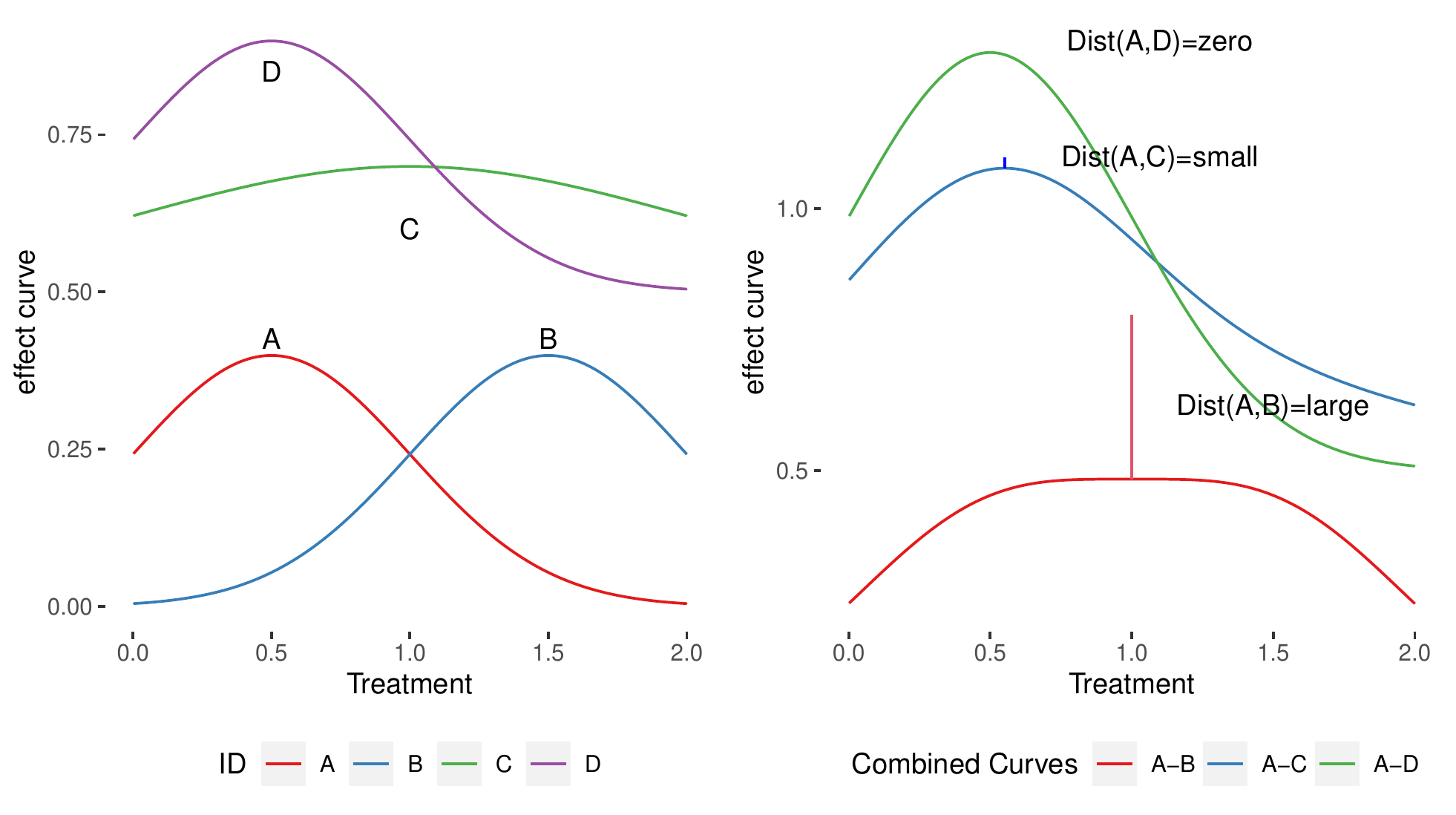}
\end{figure}

% The final similarity matrix is based on two measures, one is the effect curve shape and the other is the weighted distance based on variable importance. The ``min'' function in step 5 means, two samples are similar only when their effect curve shapes are close and their weighted distance is small. One noteworthy point is that, we used variables' main term and interaction term with treatment to fit $\hat{\mu}$, but only use the importance of interaction terms (which is the output of BART) to define $q_{p\times 1}$, because the main term effect doesn't determine the optimal treatment.)

\section{Optimal Dose Finding in Dynamic Treatment Regimes}

%The first step is to estimate the causal effect/counterfactual outcome of continuous treatment.

%\subsection{Notation}

Suppose the data $\{(\bold{X}_t,A_t,R_t)_{t=1}^T\}$ is independent and identically distributed with sample size $n$ and comes from either a randomized trial or an observational study, where $t \in \{ 1,2,\dots ,T\}$ denotes the $t^{th}$ stage, $\bold{X_t}$ denotes the patient characteristics during $t^{th}$ stage, $A_t \in \mathscr{A}_T$ denotes a bounded continuous treatment variable and $R_t$ denotes the reward of current stage following $A_t$. Let $\bold{H}_t$ denote patient history before treatment assignment $A_t$, i.e., $\bold{H}_t =  \{(\bold{X}_v,A_v,R_v)_{v=1}^{t-1},\bold{X_t} \}$. We consider the long term outcome of interest as $Y = \psi(R_1,\dots, R_T)$, where $\psi$ is a prespecified function (e.g., sum of $R_i$ or last value $R_T$).

We denote a DTR, a sequence of individualized treatment decisions, as $g = (g_1,\dots,g_T)$,
where $g_t$ maps from patient history $\bold{H}_t$ to a continuous treatment $A_t$. To define an optimal DTR, we use the counterfactual outcome framework of causal inference in a backward way. At the final stage $T$, let $Y^*(A_1,\dots, A_{T-1}, a_T )$, or $Y^*(a_T)$ for brevity, denotes the counterfactual outcome had a patient been treated with $A_T=a_T$ conditional on previous treatments $(A_1,...,A_{T-1})$, and define $Y^*(g_T):=Y^*(g_T(\bold{H}_T))$ as the counterfactual outcome under the regime $g_T$.

The performance of $g_T$ can be evaluated by the value function $V(g_T)$ \citep{QianMurphy}, which is defined as the mean counterfactual outcome had all patients followed $g_T$, i.e., $V(g_T):= E{Y^*(g_T )}$. Therefore, the optimal rule $g_{opt}$ should satisfy $V(g_{opt}) \geq V(g_T)$ for all $g_T \in \mathscr{G}_T$, where $\mathscr{G}_T$ denotes the set of all possible rules of interest. To identify the optimal DTRs, we make the standard assumptions to link the distribution law of counterfactual data with that of observational data \citep{murphy2001}.

First, we assume the consistency, i.e., the observed outcome is the same as the counterfactual outcome under the assigned treatment, i.e., $Y = Y^*(A_T)$. Secondly, we assume no unmeasured confounding assumption (NUCA), i.e., $A_T \indep \mathscr{Y}_T \mid  \bold{H}_T$, where $\mathscr{Y}_T = \{Y^*(a_T): a_T \in \mathscr{A}_T \}$ and $\indep$ denotes statistical independence. We also assume positivity of the probability density function $f_{A_T}$, i.e., $f_{A_T}(a) > \epsilon >0$, $\forall a \in \mathscr{A}_T$. Finally, we assume the continuity of the counterfactual outcome mean, i.e., $\mathbb{E}Y^*(a_T)$ is continuous about $a_T \in \mathscr{A}_T$. Under these assumptions, the optimal rule at stage T can be written as 
\begin{equation*}
    g_{T}^{opt} = \argmax_{g_T \in \mathscr{G}_T } E_{\bold{H}_{T}} 
    \Big [ E[Y\mid A_T = g_T(\bold{H}_{T}),\bold{H}_{T}] \Big].
\end{equation*}

At an intermediate stage $t$ ($1\leq t \leq T-1$), we consider $Y^*(A_1,A_2,\dots,g_t,g_{t+1}^{opt},\dots,g_{T}^{opt} )$, the counterfactual outcome under optimal rules for all future stages, had a patient following
$g_t$ at stage $t$, given $A_1,\dots,A_{t-1}$ \citep{Moodie2012}. Similarly, under the four assumptions above, the optimal rule $g^{opt}$ at stage t can be defined as
\begin{align*}
    g_t^{opt} =&
    \argmax_{g_t \in \mathscr{G}_t } E_{\bold{H}_{t}} [Y^*(A_1,A_2,\dots,g_t,g_{t+1}^{opt},\dots,g_{T}^{opt} )]\\
    =& \argmax_{g_t \in \mathscr{G}_t } E_{\bold{H}_{t}} 
    \Big [ E[\Tilde{Y}_t\mid A_t = g_t(\bold{H}_{t}),\bold{H}_{t}] \Big],
\end{align*}
where $\mathscr{G}_t$ is the set of all potential rules at stage $t$, $\Tilde{Y}_T = Y$ at stage $T$, and can be defined recursively using Bellman's optimality at an earlier stage $t$:
\begin{equation*}
    \Tilde{Y}_t = E\Big[ \Tilde{Y}_{t+1}\mid A_{t+1} = g_{t+1}^{opt}(\bold{H}_{t+1}), \bold{H}_{t+1}    \Big].
\end{equation*}

A dynamic treatment regime $g$ consists of treatment rules at all treatment stages $t\in \{1,2,\dots,T\}$. The aim
is to find the optimal treatment decision rules $g^{opt} = (g_1^{opt},\dots,g_T^{opt})$, such that when followed by the targeted patient population, the average response outcome $Y$ conditional on individual history is optimized. In the above sections we have already introduced treatment regime optimization with single stage. When it is extended to multistage scenario, for each stage t, we just need to replace the outcome $Y$ with $\Tilde{PO}_{t}(a_t) := \hat{E}(\Tilde{Y}_t\mid A_t = a_t, \bold{H}_{t})$ and replace the covariates $\bold{X}$ with patient history $\bold{H}_{t}$.

%For simplicity, in the following sections we will firstly introduce treatment regime optimization with single stage. In single stage scenario, we denote $Z_i = (\bold{X}_i,A_i,Y_i)$ as the observed data for patient $i$, where $\bold{X}_i$ is a vector of covariates, $A_i$ a continuous treatment or exposure, and $Y_i$ the outcome of interest. $Y^a$ is the counterfactual outcome when patient take the treatment $a$ and $\mu(\textbf{x},a) = \mathbb{E}[Y^a \mid \textbf{X} = \textbf{x}]$ is the conditional expectation of $Y^a$.

\section{Simulations}

For scenario 1, we use a regular setting $A_{opt} = (x_1+x_2)/2$ (\citealp{LZ}), to test different models' performance when optimal treatment is continuous. We consider generative models in which treatments are uniformly distributed on (0,1), covariates $X$ are uniformly distributed on the p-dimensional unit cube $[0,1]^p$ and $Y=u(X) - c(X,A) + Z$, $u(X) = k_p + \tau_p \sum_{j=1}^p X_j$, where $k_p$ and $\tau_p$ are chosen so that $var\{u(X)\}=5$ and $E\{u(X)\}=  - E\{\sup_a c(X,a) \}$, Z is an independent standard normal variate, and 

 $$c(x,a) \propto \frac{1}{1+10(2a-x_1-x_2)^2}.$$

For scenario 2, we consider generative model in which treatments are uniformly distributed on (0,1), covariates $X$ are uniformly distributed on the p-dimensional unit cube $[-1,1]^p$ and $Y \sim \mu(X,A) + N(0,1)$, $\mu(X,A) =  (\sum_{i=1}^{p} x_i)/p + 100(A-A_{opt})^2 $. To illustrate the global optimality of GoDoTree, we consider a simple scenario where greedy algorithms will fail: $A_{opt} = 0.75$ when $x_1x_2\geq0$ and $A_{opt} = 0.25$ when $x_1x_2 < 0$. In this case, the interaction of $x_1$ and $x_2$ decides the pattern of optimal dose. Although $x1$ and $x_2$ are informative variables, greedy algorithms will fail in this scenario since there is no purity increment when only one of them is used as the splitting rule.

For scenario 3 and 4, we consider a generative model in two-stage scenario. Two-stage treatments $A_1$ and $A_2$ are uniformly distributed on (0,1), two-stage covariates $X$ and $Z$ are both uniformly distributed on the p-dimensional unit cube $[-1,1]^p$ and two-stage outcomes are generated as $Y_1 \sim \mu_1(X,A_1) + N(0,0.1)$, $Y_2 \sim \mu_2(Z,A_2) + N(0,0.1)$, where $\mu_1(X,A_1) = (\sum_{i=1}^{p} X_i)/p + \rho_1(A_1-A_{opt,1})^2$ and $\mu_2(Z,A_2) = (\sum_{i=1}^{p} z_i)/p + \rho_2(A_2-A_{opt,2})^2$. For scenario 3, we consider linear optimal treatments which can be detected by greedy algorithm:$\rho_1=1,\rho_2=2$, $g_{opt,1}(X)= 0.5 + (X_1 + X_2)/4$ and $g_{opt,2}(H_1,Z)= 0.5 + (Y_1 + Z_1)/4$. For scenario 4, we consider a tree type optimal treatment at stage 2 which cannot be detected by greedy algorithm: $\rho_1=2, \rho_2=10$, $g_{opt,1}(X)= 0.5 + (X_1 + X_2)/4$, $g_{opt,2}(H_1,Z)= 0.2$, when $Z_1(Y_1-0.1)>0$, and $g_{opt,2}= 0.8$ otherwise. Here $H_1=(X,A_1,Y_1)$ denotes the history data before $A_1$, the optimal dynamic treatment decision rule $g_{opt}=(g_{opt,1},g_{opt,2})$ is to minimize the long term outcome $Y = Y_1 + Y_2$.

Three algorithms for optimal dose finding are compared with the proposed GoDoTree algorithm: LZ (\citealp{LZ}), CART, and random dosing. LZ uses outcome weighted learning and a greedy approach to search for the optimal tree. CART is trained by estimates of the conditional outcome mean model to predict the optimal dose. Random dosing is used as a benchmark. To make the comparison fair, all algorithms use BART as the working conditional outcome mean model, and the propensity model is assumed to be known for LZ but unknown for GoDoTree. LZ relies on a correctly specified propensity model, while GoDoTree has doubly robustness and only requires one of the two models to be correctly specified.

\begin{table}[H]
\centering
\caption{Comparison of the proposed method GoDoTree with other methods for determining the optimal continuous treatment $A_{opt} = (x_1+x_2)/2$: n = 500, p = (10,50), $E\{Y(g_{opt})\}=0${;} reported values are based on 100 Monte Carlo replications, using a test set of size 1000.}
\begin{tabular}{|l|c|c|c|c|c|c|c}
\hline
\multicolumn{2}{|c|}{\multirow{2}{*}{Scenario 1}} & \multicolumn{2}{c|}{Tree Height = 2}& \multicolumn{2}{c|}{Tree Height = 3} \\
\cline{3-6}
\multicolumn{1}{|l}{} & & $E\{Y(\hat{g}_{opt})\}$ & RMSE of $\hat{g}_{opt}$ & $E\{Y(\hat{g}_{opt})\}$ & RMSE of $\hat{g}_{opt}$\\
\hline
\multirow{4}{*}{p=10} 
& GoDoTree & 2.20 (0.19) & 0.125 (0.01) 
           & 1.81 (0.23) & 0.112 (0.01)      \\
& LZ     &  2.07 (0.27)& 0.125 (0.01)
           &  1.90 (0.23)&  0.118 (0.01) \\
& CART       & 2.22 (0.19) & 0.136 (0.01)
           & 2.07 (0.20) & 0.128 (0.01)  \\           
& Random   & 4.51 (0.09) & 0.353 (0.01)
           & 4.51 (0.09) & 0.353 (0.01)\\
            \hline
\multirow{4}{*}{p=50} 
& GoDoTree & 2.27 (0.25) & 0.138 (0.01) &
           2.06 (0.29) & 0.127 (0.02)     \\
& LZ     & 2.12 (0.25) & 0.128 (0.01) &
         2.12 (0.24) & 0.133 (0.02)     \\
& CART   & 2.32 (0.19) & 0.140 (0.01) &
          2.26 (0.23) & 0.142 (0.01)    \\
& Random & 4.52 (0.10) & 0.353 (0.01)&
         4.52 (0.10) & 0.353 (0.01)   \\
            \hline
\end{tabular}

\label{table1}
\end{table}

\begin{table}[H]
\centering
\caption{Simulation results in scenario 2, where greedy search cannot achieve global optimum: n = 500, p = (10,50), $E\{Y(g_{opt})\}=0${;} reported values are based on 100 Monte Carlo replications, using a test set of size 1000.}
\begin{tabular}{|l|c|c|c|c|c|c|c}
\hline
\multicolumn{2}{|c|}{\multirow{2}{*}{Scenario 2}} & \multicolumn{2}{c|}{Tree Height = 2}& \multicolumn{2}{c|}{Tree Height = 3} \\
\cline{3-6}
\multicolumn{1}{|l}{} & & $E\{Y(\hat{g}_{opt})\}$ & RMSE of $\hat{g}_{opt}$ & $E\{Y(\hat{g}_{opt})\}$ & RMSE of $\hat{g}_{opt}$\\
\hline
\multirow{4}{*}{p=10} 
& GoDoTree    & 2.51 (1.29) & 0.15 (0.04) & 2.35 (0.84) & 0.15 (0.02) \\
& LZ           & 7.91 (2.31) & 0.65 (0.04) &  7.00 (3.38) & 0.66 (0.04)  \\
& CART         & 5.71 (0.71) & 0.24 (0.02) & 5.26 (0.96) & 0.23 (0.02)  \\
& Random       & 14.55 (0.52) & 0.38 (0.01) & 14.55 (0.52) & 0.38 (0.01)  \\
 \hline
\multirow{4}{*}{p=50} 
& GoDoTree & 4.54 (1.19) & 0.21 (0.03) & 4.34 (1.05) & 0.21 (0.03)  \\
& LZ       &10.95 (1.85) & 0.66 (0.04)   & 8.43 (1.45) & 0.67 (0.05)  \\
& CART    & 6.24 (0.24) & 0.25 (0.004) & 6.24 (0.27) & 0.25 (0.005) \\
& Random  & 14.53 (0.54) & 0.38 (0.01)& 14.53 (0.54) & 0.38 (0.01) \\
\hline
\end{tabular}

\label{table2}
\end{table}

In Tables 1 and 2, we report the average performances and corresponding standard deviations of GoDoTree and several other methods for the single-stage scenario, using two evaluation criteria: $E\{Y(\hat{g}_{opt})\}$ (the expected counterfactual outcome when patients take the decision rule $\hat{g}_{opt}$, with an lower bound of $E\{Y(g_{opt})\}=0$ for both scenarios) and root mean square error (RMSE, the square root of $E\{g_{opt}-\hat{g}_{opt}\}^2$, measuring the difference between the proposed decision rule $\hat{g}_{opt}$ and the optimal decision rule $g_{opt}$). The results in Table 1 show that both LZ and GoDoTree outperform CART, and that the decision trees are robust with additional noise variables and varying tree height. In Table 2, we find that GoDoTree achieves the best performance among all methods: it is the only method that recognizes the global pattern (i.e., the interaction between $x_1$ and $x_2$), and thus has the highest counterfactual outcome mean and the smallest RMSE for $\hat{g}_{opt}$. In contrast, CART and LZ fail to recognize the global pattern and get trapped in local optima due to their greedy searching approach.

Table 3 shows the average performances of the above methods in two-stage scenarios, which are consistent with the results of the single-stage scenario. In scenario 3, where the optimal treatment structure is simple and can be identified by greedy algorithms, all three methods have comparable results. In scenario 4, where the optimal treatment has an underlying tree structure that cannot be detected by greedy algorithms, GoDoTree outperforms the other two methods due to its global optimality.

\begin{table}[H]
\centering
\caption{Simulation results of scenario 3 and scenario 4, which are two-stage DTRs: n = 500, p =10, tree height = 3, $E\{Y(g_{opt})\}=0${;} reported values are based on 100 Monte Carlo replications, using a test set of size 1000.}
\begin{tabular}{|c|c|c|c|c|c}
\hline
\cline{3-5}
\multicolumn{2}{|c|}{Two-stage DTR} & $100E\{Y(\hat{g}_{opt})\}$ & RMSE of $\hat{g}_{opt,1}$ & RMSE of $\hat{g}_{opt,2}$\\
\hline

\multirow{3}{*}{Scenario 3} 
& GoDoTree & 1.1 (0.6) & 0.018 (0.008) &0.006 (0.002)  \\
& LZ       & 2.5 (1.2) & 0.051 (0.015) &0.013 (0.013)  \\
& CART     & 0.9 (0.6) & 0.013 (0.003) &0.005 (0.001)  \\
 \hline
\multirow{3}{*}{Scenario 4} 
& GoDoTree    & 6.2 (1.6) & 0.019 (0.008) & 0.063 (0.015)  \\
& LZ           & 11.3 (3.7) & 0.078 (0.044) &  0.115 (0.040)\\
& CART         & 8.9 (0.8) & 0.016 (0.007) & 0.090 (0.005) \\
\hline
\end{tabular}
\label{table2}
\end{table}

\begin{figure}[H]
\caption{Comparison of decision trees generated by the proposed method GoDoTree with the truth and two other methods CART \& LZ. The first two levels of GoDoTree are nearly identical to the optimal decision rules, while there are many sub-optimal nodes in CART and LZ decision trees.}
\centering
 \includegraphics[scale=0.3]{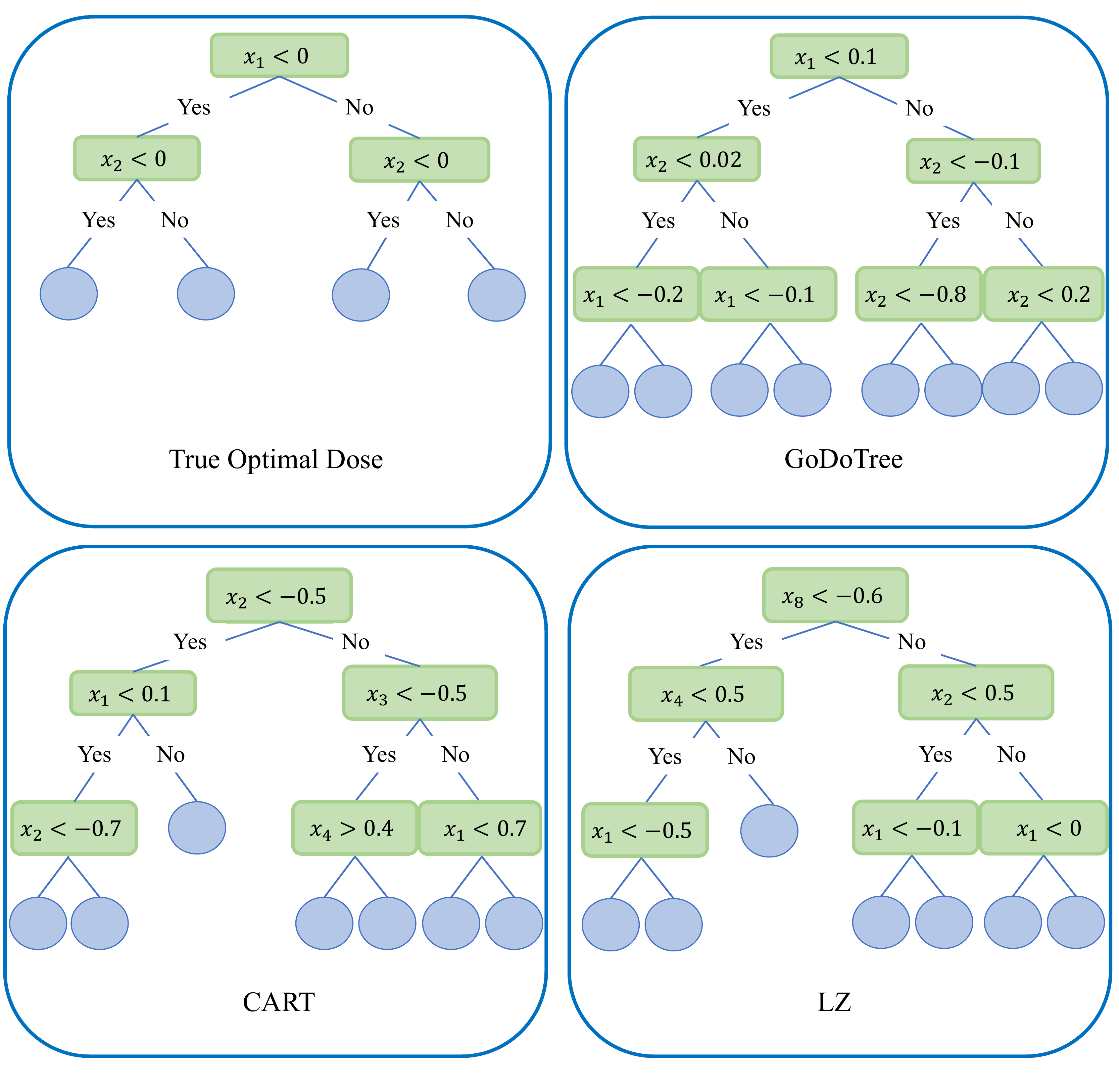}
 \label{figure3}
\end{figure}

Figure 3 highlights the limitations of greedy methods. In scenario 2, when using a single informative variable $x_1$ or $x_2$, the greedy methods are unable to select the optimal decision rule as the first splitting node, as this does not result in a higher objective function at the early stage. Consequently, the greedy trees may choose a sub-optimal decision rule as the first node, which cannot be updated later, leading to an overly complicated tree. In the CART model fitting process, there is a local optimum that is also the global optimum when $x_2\textless -0.5$, and the decision rule $x_1\textless0.1\approx 0$ is nearly optimal. However, since the first node $x_2\textless -0.5$ is sub-optimal, it is even more difficult for the second level nodes to identify the correct pattern. As a result, CART selects a sub-optimal rule $x_3\textless -0.5$ again. While increasing the tree height and size may improve the probability of detecting the correct pattern, it also increases the risk of overfitting and reduces interpretability.

%generative model in which treatments are uniformly distributed on (0,1), covariates $X$ are uniformly distributed on the p-dimensional unit cube $[-1,1]^p$ and $Y \sim \mu(X,A) + N(0,1)$, $\mu(X,A) = 100 + (\sum_{i=1}^{10} x_i)/10 - 100(A-A_{opt})^2 $. The optimal treatment assignment is $A_{opt} = 0.75$ when $x_1x_2\geq0$ and $A_{opt} = 0.25$ when $x_1x_2>0$.

% In the continuous treatment case, we consider generative models in which treatments are uniformly distributed on (0,1), covariates $X$ are uniformly distributed on the p-dimensional unit cube $[0,1]^p$ and $Y=u(X) + c(X,A) + Z$, $u(X) = k_p + \tau_p \sum_{j=1}^p x_j$, where $k_p$ and $\tau_p$ are chosen so that $var\{u(X)\}=5$ and $E\{u(X)\}= 10 - E\{\sup_a c(X,a) \}$.

%$$Model 1: c(x,a) \propto \frac{1}{1+10(2a-x_1-x_2)^2}$$

%\begin{align*}
%  Model 2: c(x,a) &\propto 1_(x_1 \geq 0.7) \phi[3\{ \Phi^{-1}(a) + \Phi^{-1}(0.75)  \} ] +
%1_(x_1 \geq 0.7,x_2 > 0.5) \phi[3 \Phi^{-1}(a) ] \\
%&+ 1_(x_1 < 0.7,x_2\leq 0.5) \phi[3\{ \Phi^{-1}(a) + %\Phi^{-1}(0.25)  \} ]   
%\end{align*}

\section{Real Application for Optimal Warfain Dose Finding }

 Warfarin is a commonly used anticoagulant medication that requires precise dosing to prevent harmful blood clots. Overdosing predisposes patients to a high risk of bleeding, while underdosing diminishes the drug’s preemptive protection against thrombosis. The international normalized ratio (INR) is used to measure how rapidly the blood can clot and is monitored to ensure that the dose of Warfarin is safe and effective. For patients prescribed Warfarin, the optimal therapeutic INR range is typically between 2 to 3 \citep{INR2to3}. To convert the INR to a direct measure of reward, the literature codes reward $R = - 100 \times \lvert INR - 2\rvert - 100 \times  \lvert INR - 3\rvert $, which is a concave function and reaches maximum between 2 to 3 \citep{warfarin_intro}. 

The dataset provided by \cite{warfarin} consists of 1780 subjects, including information on patient covariates, final therapeutic warfarin dosages, and patient outcomes (INR). The variables include weight, height, age, use of certain medications, gender, race, and genetic information. The optimal dose decision tree generated by GoDoTree is shown in Figure 4, and the results agree with well-established medical knowledge in the literature\citep{commonsense}. for example, patients with VKORC1 homozygous A/A and CYP2C9 alleles 2 or 3 requires a lower dosage of warfarin. Furthermore, our estimated results suggest to offer optimal dosage numerically, considering the possible interaction between genes and other biomarkers (race, gender, height, and weight). Our decision tree also suggests that Asian people should receive a higher dose, which may appear to contradict common sense. However, this is because our analysis only considers the criterion of ``INR located between 2 to 3'' as the optimal result, while in practice physicians may consider additional factors that could result in an optimal INR outside of this range.

 \begin{figure}[H]
\caption{Optimal Warfarin dose decision tree generated by GoDoTree, with individual and population-level estimated effect curves shown for each leaf node. Quantile information for weight and height is used due to de-identified data. Effect curves are normalized to have a maximum value of zero since only the optimal dose is of interest.}
\centering
 \includegraphics[scale=0.39]{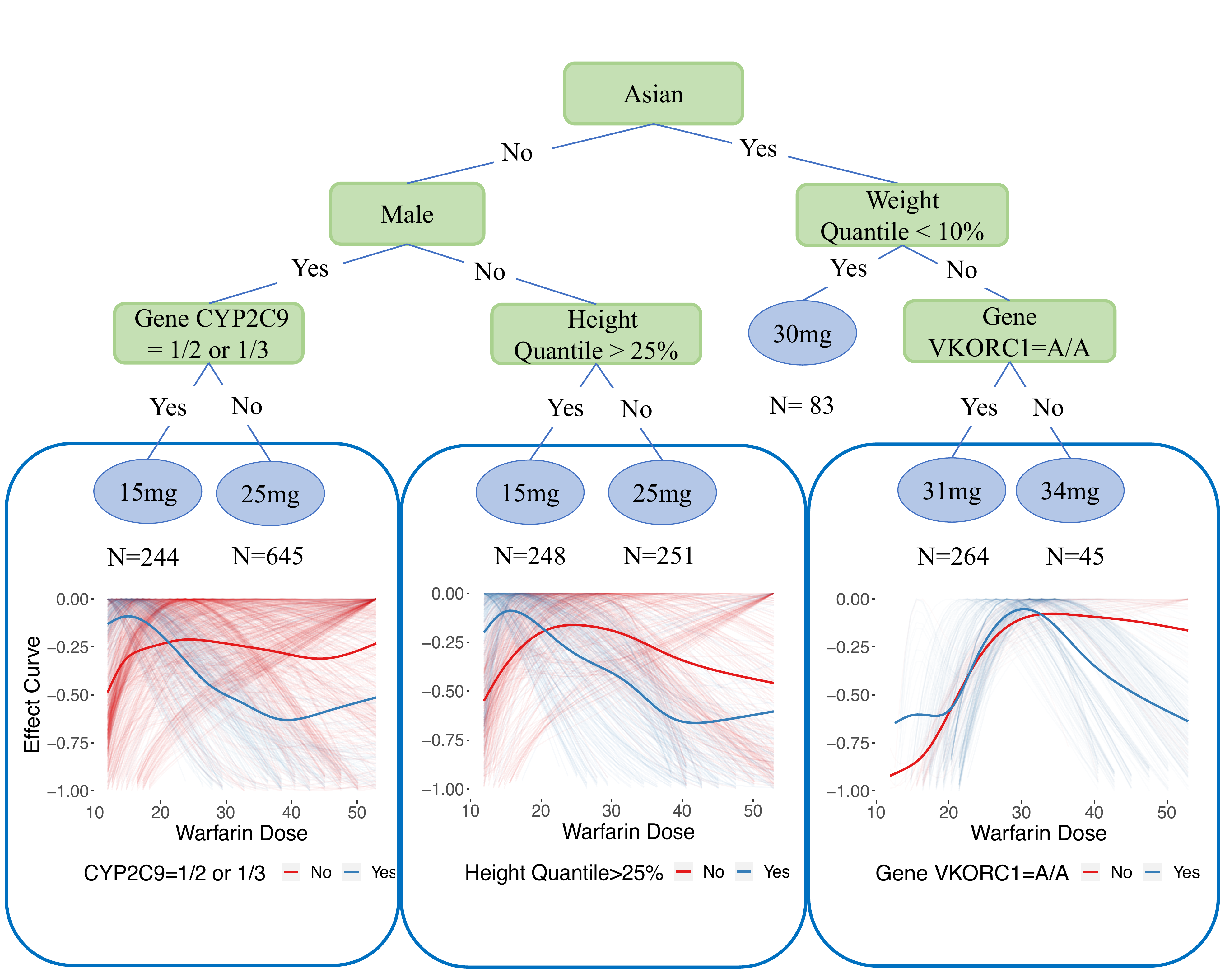}
\end{figure}

\section{Discussion}

% The proposed GoDoTree contributes to the research of dynamic treatment regimes with continuous treatment dosage and makes it possible to design a decision tree with both global optimality and great interpretability. These two major advantages are attributed to the combination of individualized causal inference and non-greedy tree searching, which enable it to update sub-optimal nodes iteratively and converge to global optimum finally. In real applications, the global optimality allows us to achieve great performance with a simple tree, rather use many sub-optimal nodes and take the risk of overfitting. The great interpretability of tree based decision rule makes it to understand and implement in real application. As for other two theoretical properties, the doubly robustness reduces the guesswork for specifying working model and make results more stable, meanwhile the asymptotic normality makes statistical inference possible. 

The proposed GoDoTree is a novel tree-based learning approach for continuous dosage finding in multiple decision stages. It falls under the category of treatment-tree algorithms, which first estimate patient-specific effect curves and then perform a supervised tree learning. This approach has two major advantages: global optimality and great interpretability. The individualized counterfactual outcome estimation and non-greedy tree search allow GoDoTree to update sub-optimal nodes iteratively and recognize underlying patterns, resulting in better global convergence. Additionally, the implementation of the tree-based decision rule is straightforward and easy to interpret for users. GoDoTree also has two ideal theoretical properties: doubly robustness, which makes the results more stable, and asymptotic normality, which makes statistical inference possible.

Due to greedy tree's lack of optimality, some previous efforts have been made to search for global optimal tree search, e.g. \cite{Hu2019} optimized sparse decision tree with penalized objective function. But most of them are supervised learning and cannot be combined with DTR learning easily because individual counterfactual outcome is missing and only population level causal effect can be estimated. There are also literature using stochastic search as the remedy, e.g., \cite{STRL} is introducing random change of tree structure with Markov Chain Monte Carlo (MCMC) to optimize the decision rule. However, although it has improvements over random search, it still lacks efficiency because of the nature of unsupervised learning. An important contribution of our work is the development of individualized counterfactual outcome estimation, which enables the global optimal tree learning of DTRs. The node updating in tree learning is a simple bi-partition task with supervision, which significantly speeds up the optimization training.

GoDoTree can be easily extended to scenarios with categorical treatment and this work mainly focus on optimal dose finding because it has not been covered thoroughly. The same approach also works when there are multiple objective functions to be optimized as long as with a well defined utility function. The optimal kernel search of GoDoTree can also be used to raise the information-noise ratio based on the estimated variable importance, making it suitable for sparse and high-dimensional data.

GoDoTree contributes to the development of precision health by enabling precise dose finding for each individual and tailoring interventions to every single patient. It can help clinicians search, validate, and refine new decision rules, especially in dose finding of new drugs and radiation oncology. The optimal dose can be quantified, rather than roughly estimated, or categorized by experience.

\bibliographystyle{apalike}
\bibliography{ref}
\end{document}